# A Survey on Personal Image Retrieval Systems

Amit K. Nath*

PhD Candidate, Florida State University, nath@cs.fsu.edu

Andy Wang

Professor, Florida State University, awang@cs.fsu.edu

The number of photographs taken worldwide is growing rapidly and steadily. While a small subset of these images is annotated and shared by users through social media platforms, due to the sheer number of images in personal photo repositories (shared or not shared), finding specific images remains challenging. This survey explores existing image retrieval techniques as well as photo-organizer applications to highlight their relative strengths in addressing this challenge.

## 1 INTRODUCTION

Photographs are persistent representations of our memories. As a result, many people take photos frequently using various devices. With the rapid advancement of modern technologies in recent years, devices like smartphones, tablets, and digital cameras have enabled users to take many pictures without worrying about depleting film reels like they used to do in the days of analog cameras. With improvements in cloud and storage technologies, users do not need to worry much about storage capacities. Most smart devices can store many images and upload images to cloud storage and social-network-based sharing platforms as needed. According to [1], approximately 1.4 trillion photos were taken in 2020.

    Personal image collections are generally accessible only to the owner and stored in the owner's local storage media. However, social networking sites such as Facebook, Twitter, Snapchat, and Instagram are popular sharing platforms. As a result, people are used to sharing their images with friends and family through these platforms. The recent and rapid development of technologies has led to a significant rise in the number of photos taken by each person. Today, most people have smartphones equipped with high-resolution front and back cameras that enable users to capture top-quality images, and smartphones have become the most popular device for taking images. The ubiquitous access to smartphones has made taking photos a lot quicker and more convenient.

    According to studies [1], the total number of photos taken in 2018 and 2019 was 1.31 and 1.42 trillion, respectively. The projected numbers of photos taken for 2020, 2021, and 2022 are 1.43, 1.44 and 1.56 trillion, which indicate 0.8%, 0.2%, and 8.3% increases, respectively. As for the number of photos stored every year, the number

of photos stored in 2019 was 6.5 trillion, and the projected numbers for 2020, 2021, and 2022 are 7.4 trillion, 8.3 trillion, and 9.3 trillion, indicating an increase of 14.2%, 12.5%, and 12%, respectively. Both the numbers of images taken, and images stored indicate a rising trend in the number of images taken and stored worldwide.

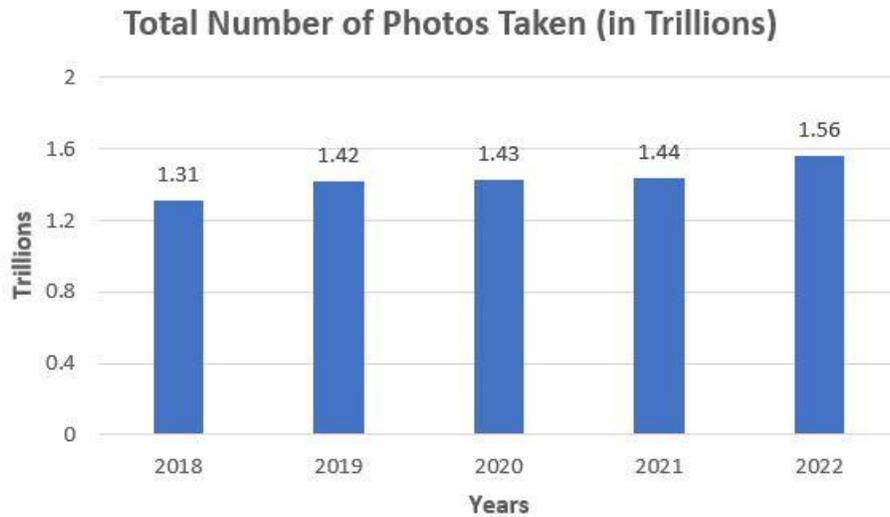

Figure 1: Increasing trend of photos taken worldwide [1].

(https://focus.mylio.com/tech-today/how-manyphotos-will-be-taken-in-2020)

However, a problem arises when a user is trying to find a specific image or image group. The search can be easily performed if the user can recall some key identifying features that are present in the image. These identifying features can include the persons or objects present in the image, landscapes, date of capture, etc. However, it is not always possible for the user to recall these identifying features. As a result, personal image retrieval remains a challenging research problem. The major challenges can be listed as follows:

**Accessing and analyzing a large personal image collection**: Due to several technological advancements, users can now store and share a huge number of images. With this trend, it is becoming very difficult to access and analyze all these images for retrieving a specific image or image group. Processing a user's whole image collection is becoming more complex and time consuming due to the ever-increasing number and resolution of photos.

**Curse of dimensionality**: A problem faced by most existing systems is the curse of dimensionality [84]. Basically, the low-level image features (e.g., color, texture, shapes, spatial layout, etc.) and user conceptual notion of image content form a high-dimensional feature space. The accuracy of categorizing images decreases as the number of dimensions increases.

**Semantic gap**: In the late 2000s, researchers started showing active interest in semantic-based image retrieval for the first time [88]. To design an efficient application for image retrieval, understanding the difference between low-level image features (e.g., hair texture) and high-level concepts (e.g., a cat) is essential. This disparity leads to the concept termed the "semantic gap" in the context of image retrieval [70]. The semantic term-based image description can be considered the highest visual information retrieval level. This task can be highly challenging [83].

**Evaluating image retrieval systems and applications**: Since the success of image retrieval systems also depends on the user's satisfaction, it can be highly subjective, and the variance of the results can be large. Hence evaluating such systems can be difficult and would require collecting user feedback and recording and analyzing their responses. As a result, human-computer interaction-based (HCI-based) evaluation is also an integral part of personal image retrieval research.

An extensive amount of research has been conducted on developing efficient image retrieval techniques. Some of the most prominent techniques include content-based image retrieval, interactive image browsing, navigational approaches, relevance feedback, automatic image annotation, etc. Also, several software applications for image organizing and retrieval have been developed.

In this survey, we initially discuss the existing research and development in the frontend (i.e., user interface) and the backend (i.e., image processing) of image retrieval systems in general, followed by a discussion on personal image retrieval systems and photo curation systems. We also analyze some popular photo organizing applications.

## 2 IMAGE RETRIEVAL

### 2.1 Basic Concepts of Image Retrieval

#### 2.1.1 Image Representation

Digital images are generally represented as pixels or colored dots on a regular display grid. Images are commonly stored in compressed raster formats such as JPG and GIF. For image retrieval systems, understanding the representation of images is quite important. Image segmentation, image features, color histograms, texture, and shapes are some of the key concepts that need to be observed and analyzed.

#### 2.1.2 Image Features and Feature Extraction

Image features are information about color, texture, or shape that are generally extracted from an image. Image features correspond to the overall description of the image contents. Images consist of low-level, local features (red, sandy) and/or high-level, global features or concepts (beaches, mountains, happy, serene). When performing image segmentation, there are two possible considerations: whole image (global features) or parts of an image (local features).

Global features are computationally simple and are generally averages across the whole image. However, they tend to lose distinction between the foreground and the background of the image and poorly reflect human understanding of images. CHROMA [47] is an example of a retrieval system that uses global image features.

Local features segment images into parts. Two schemes for this approach are regioning and tiling. The tiling approach breaks down the image into simple geometric shapes. Tiling has similar problems to global features and will possibly break up significant objects. This scheme is computationally simple and sometimes works well in practice. However, regioning breaks down the image into visually coherent areas. This scheme can identify meaningful areas and objects from an image. The drawbacks of this scheme are that it is computationally intensive and unreliable.

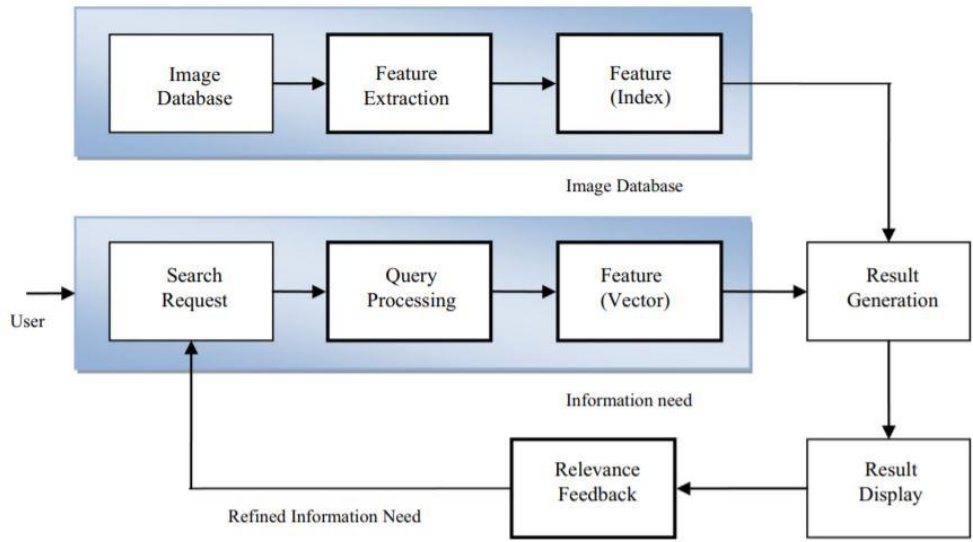

Figure 2: Overview of visual information retrieval system [71].

*2.1.3  Low-level Image Features*

- **Color**: A color signature for a region/whole image is produced using color correlograms or color histograms. There are also other color models like RGB (i.e., red, green, blue) and HSV (i.e., hue, saturation, value) used for color signatures. The problem with using this feature is that it is difficult to achieve something like human vision, since there are individual differences in human vision.
- **Texture**: Texture produces a mathematical characterization of a repeating pattern in the image (e.g., smooth, sandy, grainy, stripy). It reduces an area/region to a set of numbers that can be used as a signature for the region. Although this has proven to work well in practice, it is difficult for people to understand.
- **Shape**: Although shape belongs to the realm of object recognition, it is difficult and so less commonly used. All objects have closed boundaries and shape interacts strongly with segmentation.

These low-level features (e.g., color, texture, and shape) can be used to describe image contents individually, but the description retrieved from them is insufficient. In this context, scale-invariant feature transform (SIFT), image histograms, and CNN convolutional neural network (CNN)-based computer vision techniques are more useful for extracting informative content. In addition, feature aggregation techniques, including vectors of locally aggregated descriptors, Fisher vectors, and bags of visual words provide fixed-length vectors, which can help in approximating the performance of similarity metrics [71].

*2.1.4  Indexing Images*

Two key operations for image retrieval are image indexing and query processing. For image retrieval operations, the query speed needs to be very fast to be interactive. On the other hand, indexing or crawling speed can be comparatively slower, but indexing should still be completed as quickly as possible.

An image index is a compact data structure that supports rapid application processing. For speeding up the retrieval process, the feature-based indexing technique has proven to be useful and necessary. General principal requirements of information retrieval systems are the index size, parallelism, and the speed of index generation and search [77].

*2.1.5 Query Processing*

For a query image, the image retrieval process begins with the feature extraction operation. The first task is to extract and match the query features with pre-computed image features, taking into consideration the scalability of descriptions of visual features and the user's intent for searching. The query processing task can vary based on the indexing type being used and the features that have been extracted.

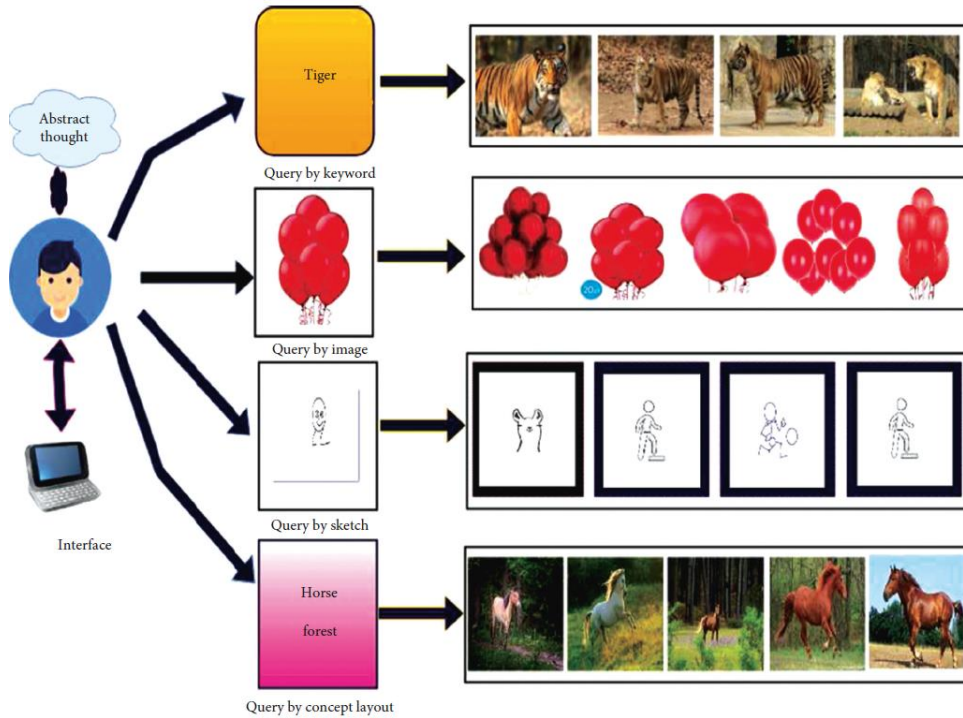

Figure 3: Illustration of different query schemes [48].

*2.1.6 Query Formation*

To define a user's subjective inquiry and requirements, query formation is a necessary endeavor. Capturing the human user's perception and intention with a query is critical but quite difficult. Several query formation schemes exist including query by text, image example, sketch, color layout, and so on, as illustrated in Figure 3.

*2.1.7 Relevance Feedback*

Various user query intents might include image quality, clarity, and associated metadata. Query refinement and iterative feedback techniques using earlier user logs and semantic feedback are highly recommended for satisfying users. The final aim is to optimize the system-user interaction during sessions. Feedback methods can be short-term techniques (modifies queries) and long-term techniques (uses query logs) [71].

## 2.2 Image Retrieval Systems

Due to the increasing number of digital images every year, image retrieval has emerged as a common operation for users, and it has thus emerged as a critical and important research area that lies at the intersection of computer vision, image processing, and human-computer interaction.

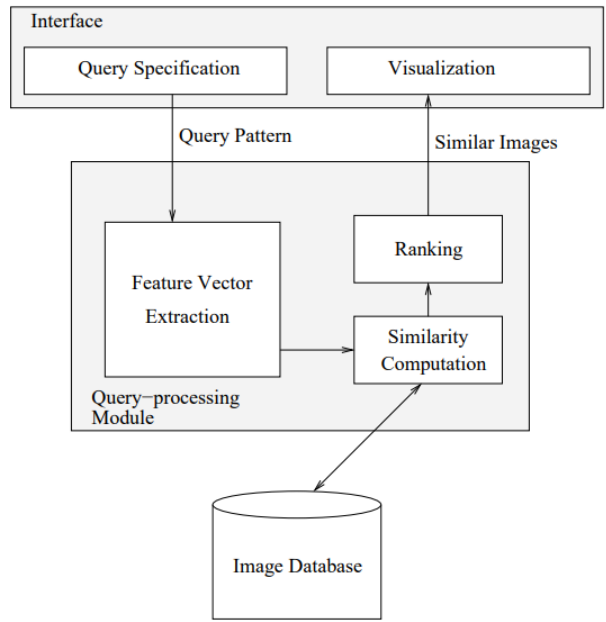

Figure 4: A typical image database retrieval system [78].

Image retrieval systems are generally computer-or smart-device-based systems used to browse, search, and retrieve images. Several image retrieval systems have been developed and explored over many years. An efficient image retrieval system needs to have the ability to search and arrange images as well as maintain a visual semantic relationship with the query provided by the user. Traditional image retrieval methods generally add metadata or annotations by using captions, keywords, titles, or descriptions. The retrieval operations are performed over annotated words and tags. Banireddy Prasaad et al. [66] developed the first microcomputer-based image database retrieval system in the 1990s at MIT. However, due to several problems, such as time and labor constraints as well as high cost, manual image annotation was not an affordable solution. As a result, significant research has been performed on automatic image annotation (AIA). Web-based image annotation tools are also being developed due to the increase in web applications and social media platforms.

Two key requirements for image retrieval systems used to explore digital image databases are ease of use and accuracy. As a result, image retrieval tools with advanced search capabilities have been identified as a necessity. In earlier times, most search engines used the text-based image retrieval technique. Most digital images to be mined were either unlabeled or inaccurately annotated. Therefore, it became necessary to perform manual annotation of the entire image collection. Performing such a manual, labor-intensive task on today's large image databases is cost-prohibitive.

Content-based image retrieval (CBIR) systems appeared as a working solution for addressing the limitations and challenges faced by text-based image retrieval systems. AIA, relevance feedback (RF), and some other techniques enhance CBIR systems. These techniques will be discussed in detail in upcoming sections.

**3 IMAGE RETRIEVAL: BACK-END RESEARCH**

A vast majority of the existing research conducted on image retrieval has focused on the back-end portion of the process. CBIR, RF, AIA, web image search and indexing are some of the key back-end retrieval techniques that have been explored at length [21]. Most of the above-mentioned back-end techniques will be discussed in this section.

**3.1     Content-based Image Retrieval**

CBIR systems emerged to overcome the limitations of the existing text-based image retrieval systems. Digital images that are mined using CBIR systems are most often represented by using a set of visual features. Figure 5 shows that typical CBIR systems generally have two phases: the offline phase and the online phase. The offline phase extracts and stores visual feature vectors retrieved from the images in the database. The online phase enables users to initiate retrieval operations by collecting the query image from the user. In the final step, CBIR systems provide a set of images that are visually similar or relevant to the given query image. This approach has a major drawback due to the initial assumption that semantic resemblance is always represented by visual similarity. The semantic gap (i.e., the gap between higher-level contextual meaning and low-level image features) is the main reason for this problem. Although Yahoo and Google achieved promising results for large-scale applications, addressing and solving the semantic gap issue remains a crucial challenge for CBIR. The increasing popularity of smart devices and social media platforms has acted as an indicator of a required paradigm shift in CBIR research. Research works emphasizing the key components of CBIR systems have been carried out, focusing on image representation, feature extraction, and similarity computation [36].

*3.1.1   Low-level Feature-based CBIR*

Several low-level features have been explored for encoding image content for CBIR systems:
- **Color feature**: For CBIR systems, color is the most popular and commonly used low-level descriptor. For color representation, various color spaces have been defined. The color spaces best perceived by humans are - RGB, LUV, HSV, YcrCb etc. For CBIR systems, multiple color descriptors and features have been explored, including color histogram, color coherence vector, color-covariance matrix, etc. Most color feature-based systems are unable to express high-level image semantics. Averaging the color of all pixels in a region as a color feature has been proposed as a solution to this problem; however, this creates another problem by affecting the image quality for subsequent processing [36].
- **Texture feature**: Texture is one of the most crucial features of an image and is widely used in pattern recognition systems. Compared to color features, texture features can be more meaningful in semantic contexts because of their ability to represent a group of pixels. The difference in texture can be useful in denoting the differences between the areas of images with similar colors. Texture can be of different categorizations. Mainly, it is categorized into two types: statistical and structural. However, the shapes of objects that are present in the image determine the semantic representation of the texture features, and these features are sensitive to noise [48].

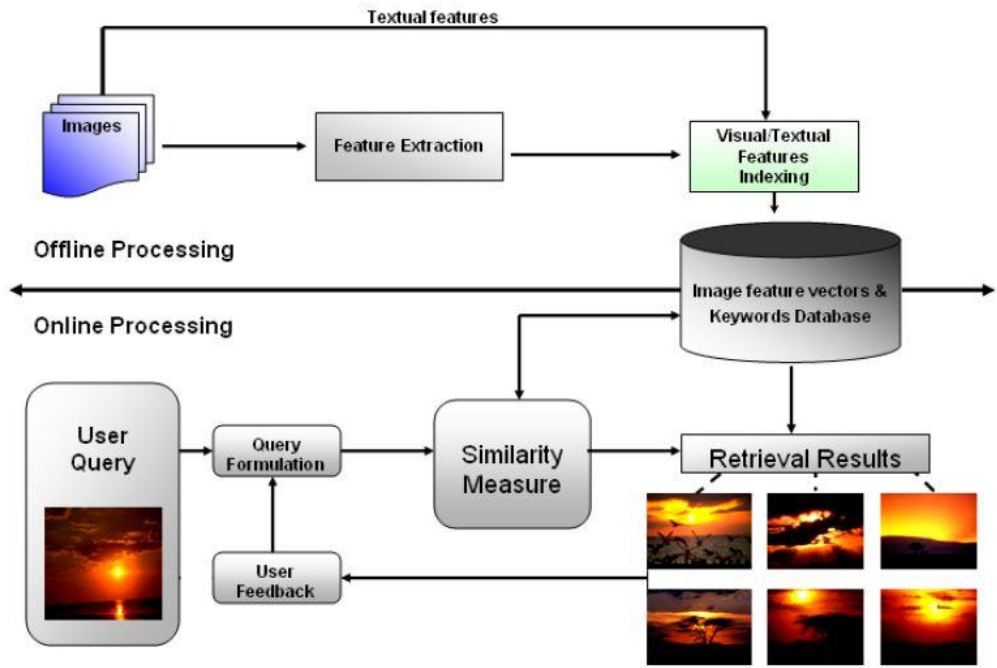

Figure 5: A typical content-based image retrieval system [36].

- **Shape feature**: Binary images for image retrieval can be achieved through shape-based query image processing. The user provides the query as a shape image, and then features are extracted from the shape and compared with the database features for constructing a CBIR system. A method was proposed in [32] that depend on shape-based queries using contour-based edge points in a binary image environment, working with rotation-invariant features. Nafaa et al. [57] developed a mechanism for a shape based CBIR system. Li et al. [50] proposed a technique for a shape-based image retrieval system, which retrieves the shape images from an MPEG7 binary image database to complete the retrieval. Based on signature histograms that are constructed from the border of the objects, [28] also provides a shape based CBIR system.
- **Spatial location**: Spatial location is an important feature for CBIR systems. Objects or regions that exhibit similar texture and color properties can be identified and represented using those features [55]. To represent spatial location, the minimum bounding box and the spatial centroid of regions were used in [51]. The problem with these approaches is that the semantic information is not represented effectively. The authors in [81] introduced an image retrieval system that placed emphasis on the spatial relationship between image contents. For image retrieval, modeling of the relationship between image objects is performed. In this approach, objects are initially detected and then the image labels are determined. After that, utilizing the binary patterns, the spatial relationships are coded. Based on the similarity between the binary patterns, a matching score is computed for performing image retrieval.

*3.1.2 Addressing the Semantic Gap Problem*

Researchers have followed several approaches to develop high-level semantic-based CBIR. Two major groups for these endeavors are supervised and unsupervised learning-based techniques and fusion-based image retrieval techniques. Generally, a single similarity measure is not enough to achieve robust image ranking with significant perceptual meaning. To address this problem, learning-based solutions can be leveraged. To speed up the image retrieval operations, image classification can play a crucial role during the pre-processing phase [60, 79]. On the other hand, for speeding up the retrieval process and enhancing visualization performance for unlabeled images, unsupervised learning can be quite useful [5, 6].

- **Supervised and unsupervised learning**: Image clustering can be used first to handle unstructured image collections; the subsequent classification techniques along with the distance metrics form the image retrieval process. Previously, most CBIR research focused on similarity metrics and feature extraction techniques. To overcome the scalability problem faced by most CBIR systems while dealing with large digital image databases, clustering and fast classification components have been identified as practical solutions, partitioning [9, 11, 38] images into homogeneous categories unsupervised. Clustering can be categorized into two major types [93]- hard clustering (elements belong to specific groups) and fuzzy clustering (elements share memberships to multiple groups). Recent clustering approaches [90] allow data instances from different clusters to be issued from different density functions. These techniques can be categorized as statistical modeling, relational, and objective-function-based paradigms. Other clustering approaches, such as spectral clustering algorithms [17], have also been proposed for grouping similar images into homogeneous clusters and then using the achieved partition information to enhance the retrieval process. Objective function optimization techniques have also been explored with popular algorithms like the K-means algorithm [53]. According to various research works [26, 62], unsupervised learning, that is, clustering techniques becomes more useful when metadata is collected along with visual descriptors. For supervised learning, Bayesian classification has been explored in several research works. Some other researchers have utilized support vector machine (SVM)-based image classification techniques. Moreover, decision tree methods such as ID3, C4.5, and CART have also been explored to predict high-level categories and associate image color features with keywords [36].
- **Multimodal fusion and retrieval**: Various image retrieval techniques relying on image and text modalities have been proposed. Some fusion techniques that can be useful for image retrieval and image annotation have also been explored. The traditional fusion approach, which requires ground truth for validating the obtained rules, focuses on learning optimal rules for fusing multiple classifier outputs. Another fusion approach formulates multi-modal fusion as a two-fold problem, which proved to be more effective than the naïve approach. This approach first performs statistical modeling of the modalities and then uses unsupervised learning to optimize the solution. The offline-based approach makes the fusion learning approaches more practical. Context-dependent fusion (CDF) is another technique that has been explored [36]. In the CDF approach, initially the training samples are grouped into homogenous context clusters by a local fusion approach.
- **Image mining for CBIR**: Image-mining-based CBIR systems execute image retrieval operations based on the similarity between images, which is defined in terms of extracted features. The optimum cluster-based image retrieval approach introduced in [41] leverages the similarity information and improves the interaction of the users with image retrieval systems. The description of images based on color characteristics and compact feature vectors representing typical color distributions is used to create the image index. The system emphasizes reducing the search time and space and creating image clusters using RGB components of color images.

### 3.2 Relevance Feedback

Most CBIR systems are constrained by two important parameters—response time and accuracy. Issues with these parameters occur due to the semantic gap between the low-level image features and the high-level human-defined concepts. The RF method can help in this regard. The RF technique is basically a supervised learning method that can improve the efficiency of information retrieval systems [59]. For RF-based systems, positive and negative feedback from the user serves as the key concept. Initially, the relevant images are retrieved and then the user is allowed to select positive or negative examples collected from the first level of the retrieved image results. Then the query list is updated, and images are retrieved from the new query [45].

The Bayesian framework feature subspace and progressive learning technique were used in [75] for RF-based image retrieval. In this case, Gaussian distribution was estimated using the positive examples, and the negative ones were used to change the ranking of retrieved images. In [22], using active learning and RF, remote sensing images (many single dates as well as time series of Earth observation scenes) are retrieved. With minimal RF rounds, this approach gets better precision criteria.

In [76], for a very large image set, a new technique called iterative RF was proposed. In each step on this system, the users are given a set of images from which they must select an image matching their query. Then, a new image set is provided, and this process is repeated. Patil et al. [61] introduced image retrieval with RF utilizing Riemannian manifolds by using positive and negative user feedback on each iteration. An adjacency matrix and its corresponding Eigen vectors were used in this case.

The proposed method from [34] used the idea of pushing the log of feedback data into traditional RF systems with the intention of combining low-level image features and high-level concepts, thus enhancing image retrieval performance. For similarity-based image retrieval, this system incorporated the concept of a soft-label SVM. Tao et al. [20] used a concept called asymmetric bagging and random subspace for RF-based image retrieval. In this case, three types of SVM techniques were used. [30] employed user feedback to reduce feature size and combined feature weight refinement with queries for attaining feature adoptive RF (FA-RF). FA-RF can map the feature space with the user selection log accordingly. An RF-based image retrieval system was used in [15] to improve accuracy and time by leveraging the dynamic hierarchical semantic network. This network was established by allowing dynamic iteration levels with an RF mechanism.

However, due to the additional workload imposed upon the user, RF systems have limited commercial use despite providing effective results for image retrieval operations.

### 3.3 Automatic Image Annotation

Conventional image annotation systems generally follow a manual approach that is very time consuming and inefficient. Due to these limitations, several researchers have explored AIA techniques. One of the key motivating works for approaching the task of AIA using weak supervision or fully automatic methods is the word co-occurrence model [33]. AIA systems focus on minimizing the semantic gap between low-level image features and high-level semantic labels. Exploring various correlations between images and corresponding labels (e.g., label-label, image-image, and image-label correlation), the AIA systems learn high-semantic labels from low-level visual features.

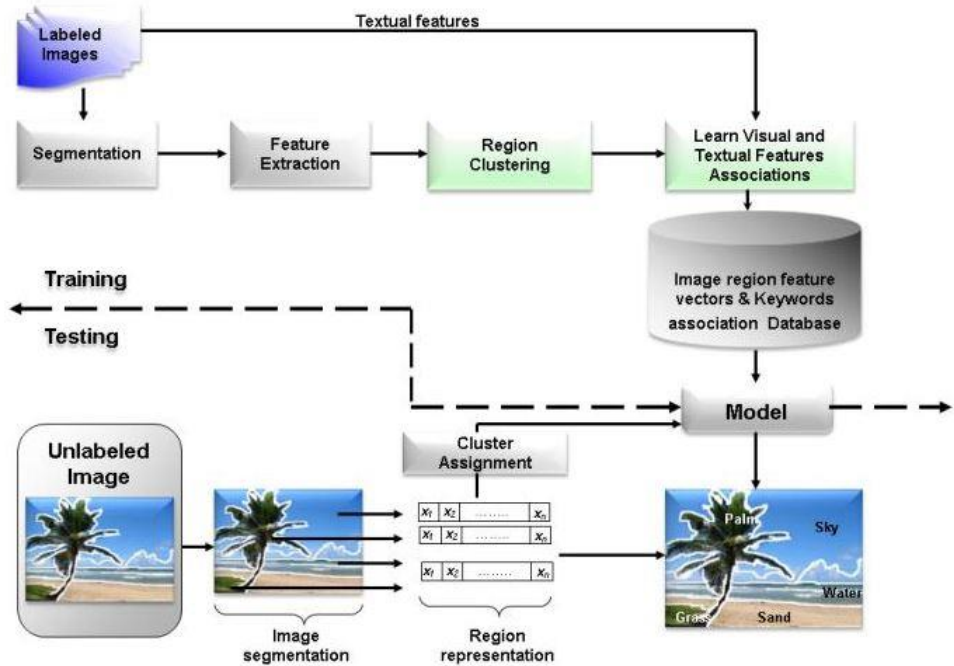

Figure 6: Structure of an image annotation system [36].

Figure 6 illustrates the architecture of a general image annotation system. A set of labeled images is used by the image annotation for training purposes. Initially, regions are created by segmenting an image. The extracted results from this process consist of local features that are useful for describing each region.

There are various classification systems for AIA, such as probabilistic and non-probabilistic techniques, methods focused on learning and retrieval, and supervised, semi-supervised, and un-supervised techniques. Such approaches can be organized into five main categories [18] –

- **Generative model-based AIA**: This method aims to maximize the generative likelihood (probability of an image label for an untagged image) of image features and image labels.
- **Nearest neighbor model-based AIA**: In this method, images having similar features possess a higher probability of sharing similar labels.
- **Discriminative model-based AIA**: In this method, the image annotation task is considered a multi-label classification problem.
- **Tag completion model-based AIA**: This method not only predict labels by filling up missing labels automatically but can also correct noisy image tags.
- **Deep-learning-based AIA**: For large-scale AIA, to derive robust visual features of exhaustive side information, deep-learning algorithms are used in this method.

Based on their key concepts, each of these categories can be divided into several sub-categories, as illustrated in Figure 7.

Researchers have been trying to develop highly accurate AIA systems. However, most of the proposed systems have significant limitations in labeling real-world images. One of the key challenges for AIA systems is efficiently annotating a massive number of web images. Using a novel feature selection method, Ma et al. [52] proposed an annotation model. The canonical correlation analysis technique was used by Gong et al. [29] for mapping textual and visual features on the same latent space and incorporating the third view for capturing high-level image semantics.

From the discussion, it is evident that a significant number of diverse learning techniques and methods have been explored for AIA, which indicates the complex and challenging nature of AIA. Overall, we can conclude that there has been extensive research work investigating the backend of image retrieval systems. However, there is still room for improvement, mostly in integrating the feedback from the user to make the systems more efficient.

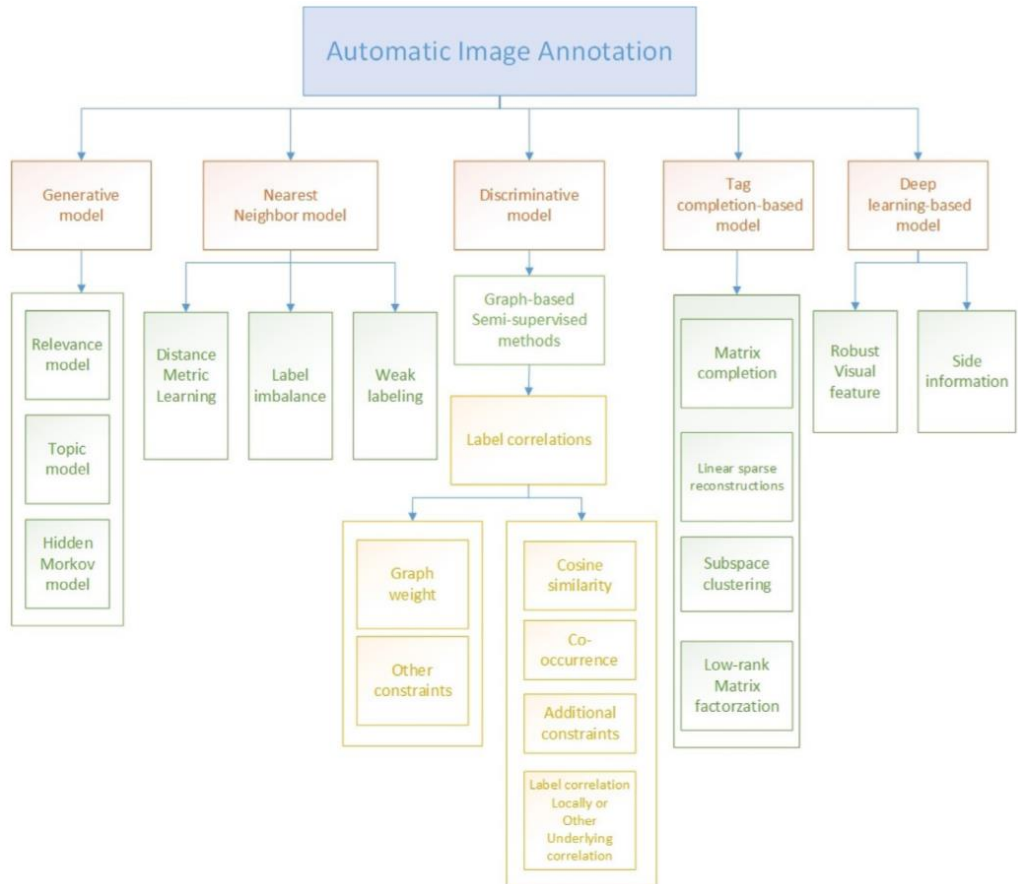

Figure 7: Categories of automatic image annotation systems [18].

## 4  IMAGE RETRIEVAL: FRONT-END RESEARCH

With the rapid growth of image databases, smart and efficient techniques to manage and query large image collections are currently very much in demand. In the previous section, we discussed techniques that mostly emphasize the back-end research of image retrieval. However, due to the emergence of touch-based smart devices, such as smartphones, tablets, and touch screens, lately there have been some front-end research works on image retrieval.

**4.1    Similarity-based Visualization**

A visual overview of the entire image database is provided in most image browsing systems, which are coupled with the required operations for navigating the images of interest. Several research works have proposed image browsing techniques as an effective alternative to back-end based image retrieval systems such as CBIR. Generally, CBIR techniques are useful for users who have clear and specific goals regarding their search items, while similarity-based image browsing is more useful for those focusing on surfing or browsing image collections. A major challenge in similarity-based image browsing is arranging the images based on their visual similarities [74].

Researchers have proposed several approaches for browsing images. In [78], the authors used spiral-based and concentric-based representation techniques for displaying similar images, keeping images with more similarity closer to the center. Visualizing an image database by linking similar images using pathfinder networks was proposed in [16]. Based on the viewpoints used for capturing photos, community photos were arranged in [73]. In [74], the authors proposed a different kind of approach by using a dynamically generated photo collage to visualize an image collection. Following the user interactions, an automatic selection of images is performed for composing the photo collage.

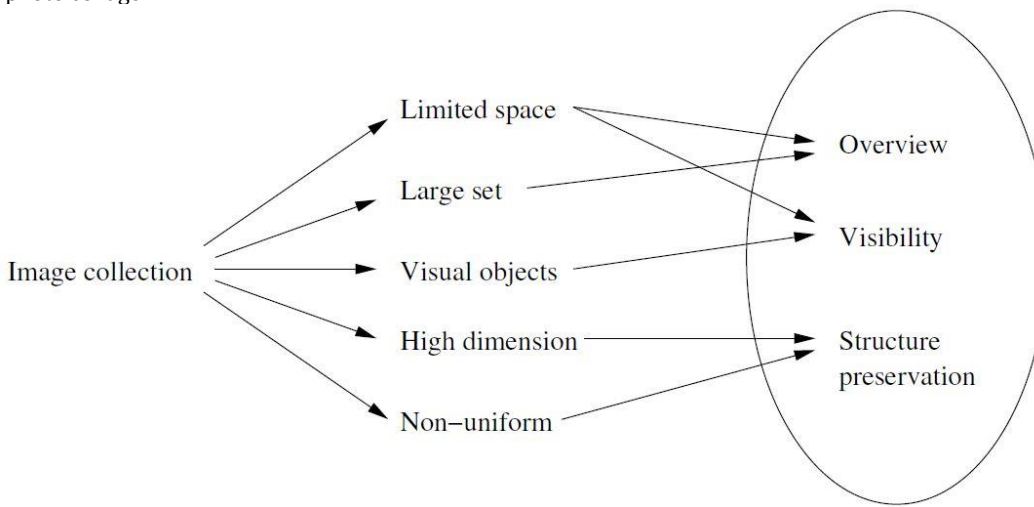

Figure 8: General requirements for visualizing image collections [58].

Some popularly used visualization schemes for large image collections are as follows [83] –
- Relevance-ordered (Google Images)
- Time-ordered (Timeline and Time Quilt)
- Clustered (Gallery layout with multi-dimensional scaling)
- Hierarchical (Google Image Swirl)
- Composite (Mix of two or more of the above-mentioned ones)

Three modes for visualization in terms of user presentation are –
- Static (No motion involved)
- Moving (Constant motion)

- Interactive (Motion triggered during user interaction)

Considering recognition success and user preference, some researchers stated that static style visualization proves to be more suitable compared to moving presentation.

### 4.1.1 Major Visualization Approaches

Three major approaches for generating visualizations of image repositories are mapping-based, clustering-based, and graph-based visualizations [64].

- **Mapping-based visualization**: In this approach, dimensionality reduction techniques, such as principal component analysis or multi-dimensional scaling, are used to preserve the relationship between images in the high-dimensional feature space in the reduced two dimensions of a computer screen.
- **Clustering-based visualization**: This approach groups similar images together and thus reduces the number of images that need to be displayed simultaneously. In this approach, several techniques can be used to define similarity, such as content-based features and timestamp data. Clustering operations are generally done using hierarchical clustering algorithms.
- **Graph-based visualization**: In this visualization technique, a graph structure is used to embed the image collection. In the graph structure, edges link related images, and nodes represent the images. Various means such as visual similarity between images and shared keyword annotations, can be used to define edges.

## 4.2 Interactive Image Retrieval

### 4.2.1 DynamicMaps

In [44], the authors presented a novel method for browsing based on similarity across a very large collection of images. The images are positioned dynamically in a high-dimensional feature space next to their nearest neighbors on a 2D canvas. During user interaction, the layout and choice of images is created on-the-fly, representing the navigation tendencies and interests of the user. A user performs imaging browsing by navigating through an infinite 2D grid, where each image's neighbors are ordered by similarity. The image map assumes precomputed k-nearest neighbors and similarity scores and is generated dynamically using localized knowledge. Updates can be performed online. Thus, for large and dynamically changing datasets consisting of millions of images, the technique is a viable solution.

Evaluation of this approach revealed that users viewed far more images per minute while using DynamicMaps compared with a traditional RF interface, indicating that it facilitates more fluid and natural interaction that allows for easier and quicker movement in the image space. DynamicMaps was favored by most users, suggesting that it is more exploratory, facilitates browsing better and is more pleasant to use.

### 4.2.2 ImgSEE

In [25], a prototype program called ImgSEE was developed by the authors to support interactive image retrieval for social media images. By using image processing and deep learning methods, they extracted new information from image, augmented the textual content of social media posts, and created a visual interface to enable interactive image retrieval.

*4.2.3    Interactive Image Browsers*

Dataset visualization can be helpful to obtain an overview of an entire image collection. However, in order for users to explore the image database and eventually arrive at the desired image, user interaction with the image database is required [63]. To this end, a distinction can be made between horizontal browsing (navigation within a single plane of visualized images) and vertical browsing (allowing navigation from one level of a hierarchical browsing structure to another).

In [69], the author developed the hue sphere image browser and the honeycomb image browser as efficient and effective methods for navigating image databases. For these methods, complex data processing algorithms are not needed, and the images are simply arranged based on their color, not considering color histograms or other complex color features. These browsers work based on the median color in HLS (hue, saturation, and luminance) color space in which the hue and lightness components are retained.

- **Hue sphere image browser**: Each image is described by two color coordinates, which are then mapped to longitude and latitude values to yield a spherical visualization. This, combined with the grid layout and hierarchical organization employed in the multi-dimensional scaling grid, leads to an intuitive and efficient approach to visualize image databases.
- **Honeycomb image browser**: Several features of the hue sphere browser are shared by the honeycomb image browser. Here also, the median image color is employed, and a regular lattice is used for placing the images. This lattice comprises space-filling hexagons for which each image corresponds to a cluster.

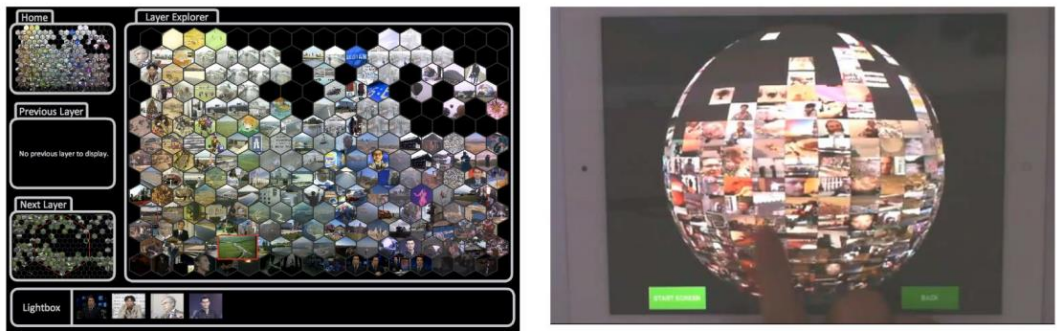

Figure 9: Honeycomb image browser and hue sphere image browser [69].

## 4.3    Applications

There are several existing smart device-based applications that can help in curating, organizing, and retrieving images. These applications are developed for android and/or iPhones. Some of the popular applications will be discussed briefly in this section.

*4.3.1    Google Photos and Apple's Photo App.*

The leading photo organizing applications for Android and iOS are Google photos and Apple's photo app, respectively. Both apps share some similar features that are very helpful for organizing and retrieving images [2].

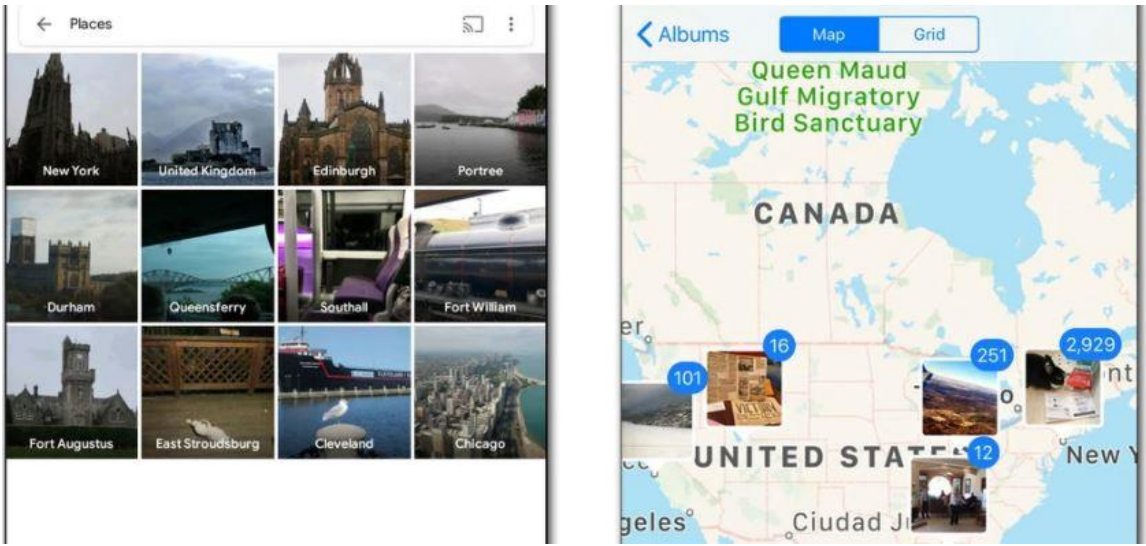
Figure 10: Location based photo album organization in Google Photos (left) and Apple's Photos app (right) [2].

- **Location-based photo albums**: If a user has the location service of the smart device activated while taking pictures, through the "geotagging" feature, the geographical location coordinates are embedded into the image file. This feature allows both Google photos and Apple's photo app to organize the photos into location-based albums, which is quite helpful for the users for retrieving images. Also, if the GPS information is not available in the photo file, locations can be added manually.

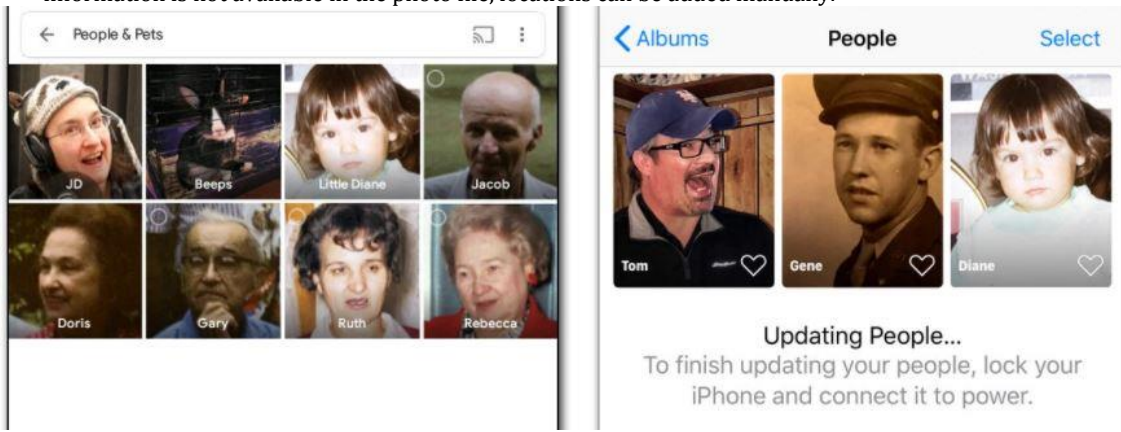
Figure 11: Grouping by automatic face scanning in Google Photos (left) and Apple's Photos app (right), the user needs to add or confirm names [2].

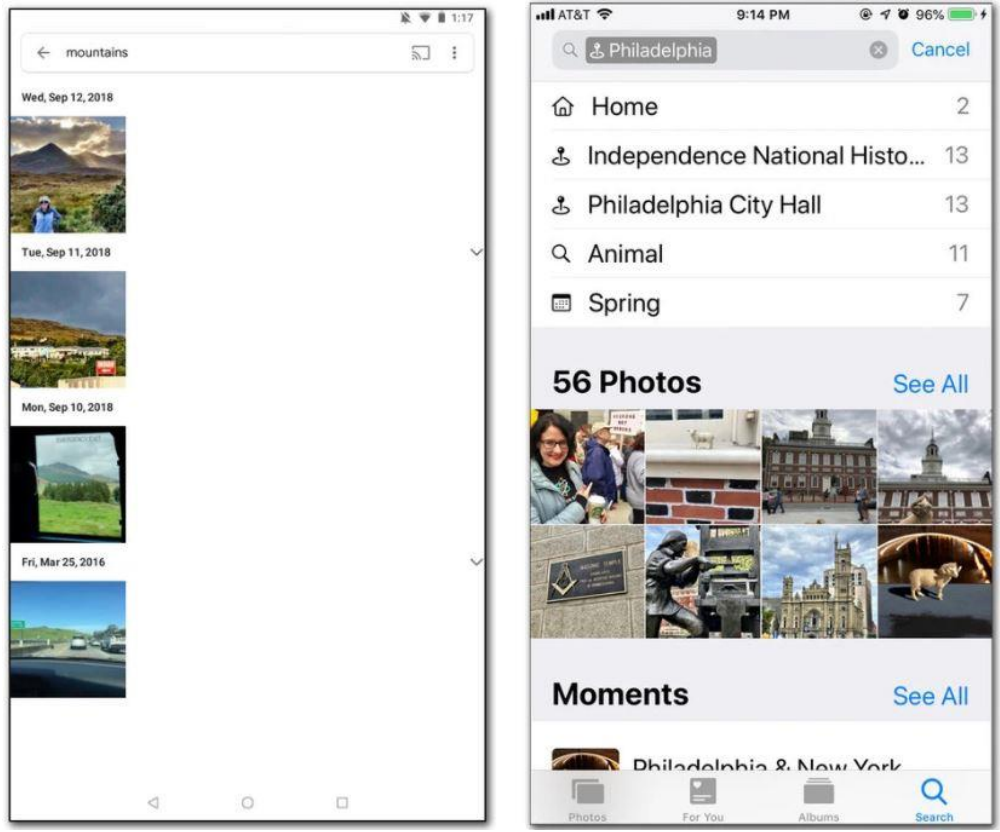

Figure 12: Search boxes can be used to find photos in both Google Photos (left) and Apple's Photos app (right) [2].

- **Person and people albums**: Unless the feature is manually disabled, similar faces are automatically grouped together by Google photos using embedded facial recognition software. These groups are shown in the "People & Pets" tab of the albums. Similarly, in iOS devices, similar faces are gathered from the picture library and displayed in the "People" area of the albums. It provides users with an option to label these groups.
- **Keyword search for finding photos**: Both programs have a search box where the user can search people, places, or things by typing keywords, or the user can also look up photos based on the timestamp (year, month, or date). If images are already tagged with the names of people and geotagged, then the search box can be used for rounding up results, like "John in Tampa in 2020." Due to the basic object recognition features provided by both apps, users can also search for items (e.g., "wedding dress" or "concerts") within images without even adding or confirming tags by themselves. Keywords such as "mountains" or "ocean" also correspond to outdoor shots.

*4.3.2   Other popular applications*

There are several other popular image organizing applications [4] available, such as Curator, Imaganize, A+ gallery, QuickPic, Piktures etc. Most of these applications have some common features like search by different criteria (date, location, image color, etc.), tagging, editing etc. Some applications have a few additional features (e.g., the Piktures app can perform optical character recognition to extract text from any photo by tapping on the option). Based on the important features, a tabular representation of these popular applications is as follows:

Table 1: Smartphone-based Photo Organizing Applications

| Applications | Keyword Search, Folder-based, Object Recognition | GPS-based | Timestamp-based | Tagging | Cloud-support |
|---|---|---|---|---|---|
| Google Photos, Apple's Photos | √ | √ | √ | Semi-automatic | √ |
| Curator, Imaganize | √ | ------- | ------- | Manual | ------- |
| A+ Gallery, QuickPic, Piktures, 3Q Album | √ | √ | √ | Manual | √ |

Overall, the front-end of image retrieval is now receiving more attention now than it used to. Although there are several existing applications and systems, most of them still have some limitations in terms of making the systems more comfortable and efficient for the users.

**5      PERSONAL IMAGE RETRIEVAL AND PHOTO CURATION**

Most existing image retrieval systems emphasize retrieving images from the internet or a large image database based on user queries. However, with the ever-increasing storage capacity in personal devices (e.g., desktop, laptop, smartphones, tablets) as well as uploaded images on cloud storage and social media platforms, each user can now have a huge number of personal images. If we consider all these sources of personal images as a single image database, it is quite difficult for the user to retrieve a specific image or set of images from this huge collection. Consequently, this has spurred the development of efficient systems that can help users in this regard.

**5.1   Human Factors in Image Retrieval**

Since individuals generate, access, and use photographs, image retrieval can be termed a human-centered process. Therefore, it is crucial to consider the conditions that correspond to both the indexing and retrieval of image content when designing and developing image retrieval systems and algorithms or when evaluating their efficiency. This involves analyzing the various retrieval interpretation levels, potential search strategies, and uses of the image. In addition, various degrees of similarity and the role of human factors (e.g., culture, memory, and personal context) must be considered [37].

While most popular search engines provide image search options, these features are focused only on keyword searches, and indexing is achieved mostly by automatically analyzing the metadata of the images (file name, URL, and surrounding text). On the one hand, because most images on the web do not have formal metadata to define their content, the approach is unstructured; on the other hand, the textual information used to index the images is frequently incorrect and incomplete. Despite these problems, for some types of searches, image retrieval using keywords and automatically indexed metadata has proven successful, especially when the metadata used describes the content accurately at the desired level. Clearly, retrieval effectiveness depends not only on the definition of the metadata, but also on how the query is conducted by the user, his expectations, and other variables.

The authors in [68] claim that the user's interaction with an image collection creates the meaning of an image for that user. The definition levels that are applicable to a specific scenario seem to depend on the collection itself, as well as on the specific query that the consumer is formulating at a given time. There are variations within search behaviors and image usage for each user. It is uncommon to have annotated pictures in personal image collections, for example, and it is difficult to scan for content because people are often more interested in finding pictures of particular people or events (e.g., pictures of a friend at another friend's birthday party). This means that it can be more effective to use browsing techniques or structures that allow efficient use of landmarks (e.g., time structure).

With the growth of personal image collection and shared images (images made public from private image collections), the scope for searching and viewing image content is also increasing. The ability to create comments, tag, and annotate images leaves the doors open for integrating visual elements, textual retrieval, and browsing for many indexing and search opportunities. Developing and evaluating adaptable systems for users could be quite challenging since the following need to be considered: human memory, context, and subjectivity. In order to develop excellent personal image retrieval systems, a holistic approach to designing and developing applications must be taken considering all the related variables concerning the final human users of these systems.

**5.2 Personalized Image Retrieval**

A big obstacle for conventional image recovery methods is enabling users to retrieve the images they need quickly and accurately. Thus, personalized image retrieval has emerged as a new trend in image retrieval that increases not only the accuracy of existing recovery systems, but also suits the needs of the users better.

*5.2.1 Personalized CBIR*

In [12], the authors developed a system to address the problem of subjectivity in CBIR systems by enabling the users to define an indexing vocabulary by themselves and making the system learn this user-defined vocabulary. On both local and global levels (object and image categories. respectively), these indexing techniques can be used. Local concepts and low-level features contribute to building global concepts making the concept learning process incremental and hierarchical. To emphasize relevant features for a specific concept, a similarity measure tuning system is used.

*5.2.2 Personalized Mulltimodal Image Retrieval*

Developed by the US Marine Corps, Quickset [19] is the first multimodal interaction-based approach for mobile systems. Combining user speech and drawing for multimodal command-based search queries, Speak4it [23] local search application is another such example. A client-server architecture was proposed in [86] for mobile devices to support multimodal search. A client-server based visual multi-modal system was also proposed in [46], while [8] proposed a system with multimodal and multi-touch functionalities with query and speech input. MAMI [7] was developed as a mobile phone prototype allowing users to annotate and search for digital photos on their phone using speech input. Not only is speech annotation at the time of image capture enabled, but the storage of additional metadata (location, time) is also available.

Given its huge volume of data, diversity of content, heterogeneous patterns of individual use and resource constraints, the management of personal image data on mobile platforms is challenging. [90] introduced a user-centric framework, iScope, to handle and share photos individually on mobile devices. Content and context information-based multi-modality clustering is used by iScope to provide efficient image management and search facilities. User-centric search algorithms, along with adaptive prediction designed for individual users, are also used in this system. Online learning techniques are used by iScope to predict images according to the users' interests.

Emphasizing energy efficiency as a primary design goal, the creators of iScope were able to improve search time and search energy compared with browsing.

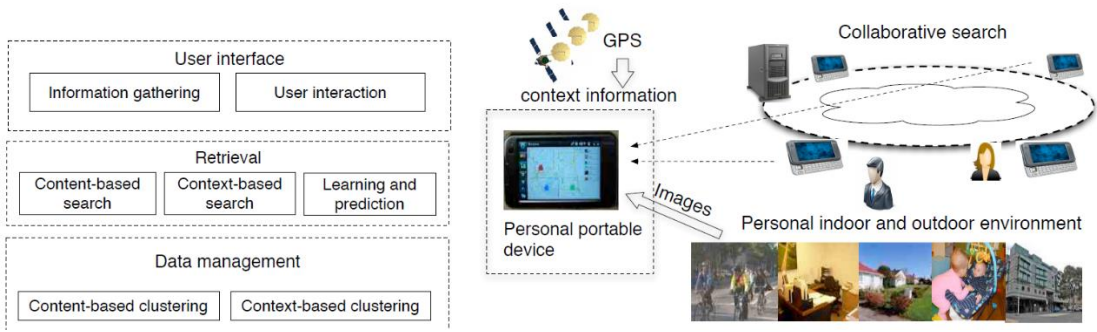

Figure 13: System architecture of iScope [91].

[87] introduced a concept termed visual-perception-based personalized image retrieval, focusing on reducing the semantic gap through direct perception of the visual information of the user. For segmenting image regions, a visual attention model is used, and record fixations are addressed using eye tracking techniques. By examining fixations in regions, visual perception is obtained to assess gaze interests. Regions of interest (ROIs) are detected through the integration of visual perception into the attention model. The features of the ROIs are extracted and analyzed and then feedback is provided as interest for constructing user profiles and optimizing results.

Another approach to incorporating visual perception into personalized image retrieval is considering user perception across modalities. The gaze of the user (way of looking at an image) can affect the image captioning (way of describing the image). In [56], the authors proposed an approach for modeling cross modality personalized image retrieval. In this system, user personality is modeled along with gaze and captions. Combining the embeddings from content and style modeling, the proposed system provides a novel approach with better results.

*5.2.3 Personalized Image Retrieval and Recommendation (PIRR)*

Personalization techniques are applied to image recovery systems to solve the issue of image overloading and increase query performance. To improve the precision of the retrieval, user preferences and image processing technologies are combined. Personalized image retrieval gets pictures passively. An image search engine helps the user locate photos. Users' inputs and personalized information drive the system to collect the target images. Relevant images are pushed to the user. As an emerging sub-field of image retrieval, PIRR is getting quite a lot of attention [27].

PIRR can be broadly classified into the following categories [39]:

- **Content-based PIRR**: The key ideas of this system are acquisition and representation of user interest and implementation of personalization. User-labeled image tags and the users' past behavior (browsing, clicking, saving, and querying) information are collected by the system. Feature extraction and semantic correlation help to deduce the users' preferences. Using this information, a user preference model is developed, which is then applied for personalized image retrieval to acquire customized results. Simultaneously, feedback from the user is collected to optimize the results.
- **Collaborative Filtering (CF)-based PIRR**: Apart from sharing information, social networking platforms may also connect users. Users typically enter a certain group in a social network to create relationships with other users who share their interests. Using the CF method, the images preferred by the user can be inferred through mining the relevance data between the users and images. User or image similarities are calculated using an image rating matrix. Heuristic methods or probability statistical methods are employed to obtain a customized search result to infer unknown ratings based on existing ratings.

While CF-based PIRR solves some of the limitations of CBIR systems using the capability to adapt to any image type, it still has some problems because it is not able to discover how the user's interest might change over time. Combining content-based systems and PIRR methods, some hybrid methods have been developed to address this problem. Efficiently embedding the content features of images into PIRR systems, protecting the privacy of the users, capturing the constant changes in user interest, etc. remain challenges for PIRR systems.

*5.2.4   User-interest-model-based Personalized Image Retrieval*

In [42], the authors introduced a new framework design called the personalized image retrieval system (PIRS), which emphasizes the preferences and interests of the user. Two major modules formed the core of the system: the user's preference module (UPM) and the retrieval module. A decision tree structure was used to implement the UPM module. The system allows the user to execute iterative retrieval activities, which enables the UPM to be repeatedly revised according to the preference of the user; thus, the PIRS can adapt to the preferences and interests of the user.

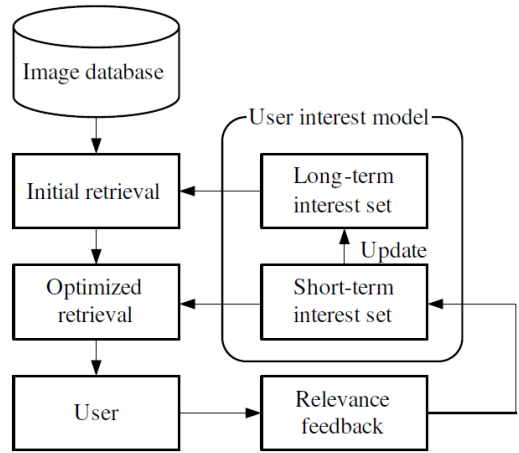

Figure 14: Structure of a user-interest-based model [40].

The user interest model plays a crucial role in reducing the semantic gap. In [40], a user-interest-model-based personalized image retrieval system is proposed. For this system, short-term and long-term interest-based user models are developed where the short-term ones are represented through a collection of visual and semantic features, and the long-term ones are inferred through inference engines from the previously gathered short-term interests.

*5.2.5 Personalized Image Retrieval in Social Media*

With the immense popularity of social media platforms, image sharing has become one of the most common activities performed by users online. Users upload various types of images, including personal photos, family photos, photos of tours, pets, celebrities, memes, etc. Generally, these images have some texts associated with them as captions that are written and posted by the users themselves. For personalized image retrieval, these shared images are crucial and need to be considered important since they mark significant events in the lives of the users and can be a good source of information for retrieving relevant images quickly.

Using a short query, people frequently try to locate an image, and images are usually indexed using short annotations. When little text is available, matching the query vocabulary with the indexing vocabulary is a difficult task. Content created by textual users on Web 2.0 platforms contains a wealth of data that can help solve this issue. In [65], the authors explained how to use content from Wikipedia and Flickr to make this match better. A generic social-media-driven query expansion model was introduced by the authors and tested on a large-scale, noisy image collection. In Flickr, the initial query was initiated, and a query model was generated based on the co-occurring terms. Using Wikipedia, the nearby definitions were also measured, and these were used to extend the query. The final results were obtained using the similarities between their annotation and the Flickr model by rating the outcomes for the extended question. The evaluation of this expansion and ranking method over the ImageCLEF 2010 Wikipedia Set, which included 237,434 images and their textual annotations, showed that compared to existing methods, a consistent improvement was achieved.

Social media platforms provide an opportunity to explore collective community behavior with analysis of the linked multi-modal data such as images and tags. Contextual information can be collected from the tags, and visual contents are represented by the images. By exploring latent feature space between visual features and context, another social image retrieval approach is proposed in [49], where context regularization terms are imposed for constraining visual features.

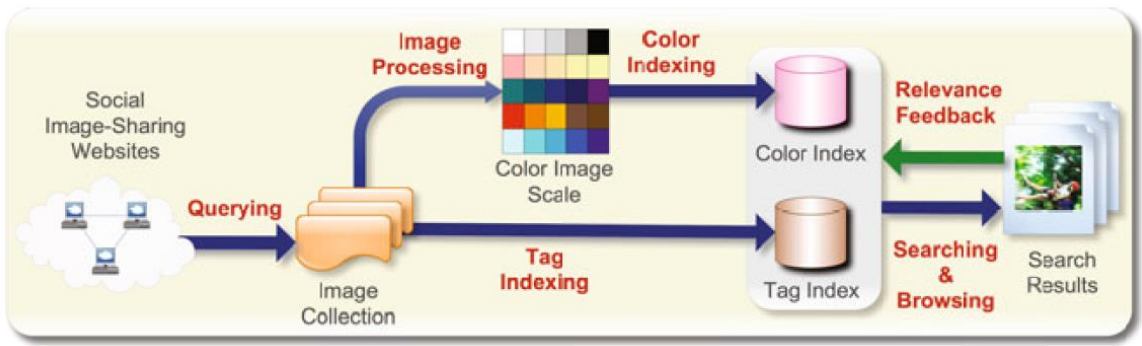

Figure 15: CBIR approach for improving tag-based social image retrieval [31].

Social media image tags can be leveraged for keyword-based image retrieval through the construction of an index from user-assigned image tags. However, tag spamming and subjectivity create some drawbacks for such systems. In [31], a CBIR approach is proposed for social media image retrieval that can improve the search results provided by tag-based image retrieval systems. In this approach, using color and tone information from an image collection, an index is constructed. The newly constructed color and tone index is used to filter the results of the keyword-based query and rank the search results using the RF technique.

[3] has provided an RF algorithm that can adapt the response of a content-based image retrieval system to the information needs of a social media user. The algorithm estimates the significance of each content descriptor to the similarity measure of the system, which helps in maximizing the correlation between the query image and all images marked as relevant by the user. To handle multiple feedback iterations, a recursive algorithm is also provided.

In [13], the authors utilized social annotations and considered user interest along with query relevance while proposing a three-step framework for personalized social image retrieval. According to user interest, the framework generates a return list from which the user-provided metadata is leveraged for personalizing the search results. The framework is illustrated in Figure 16.

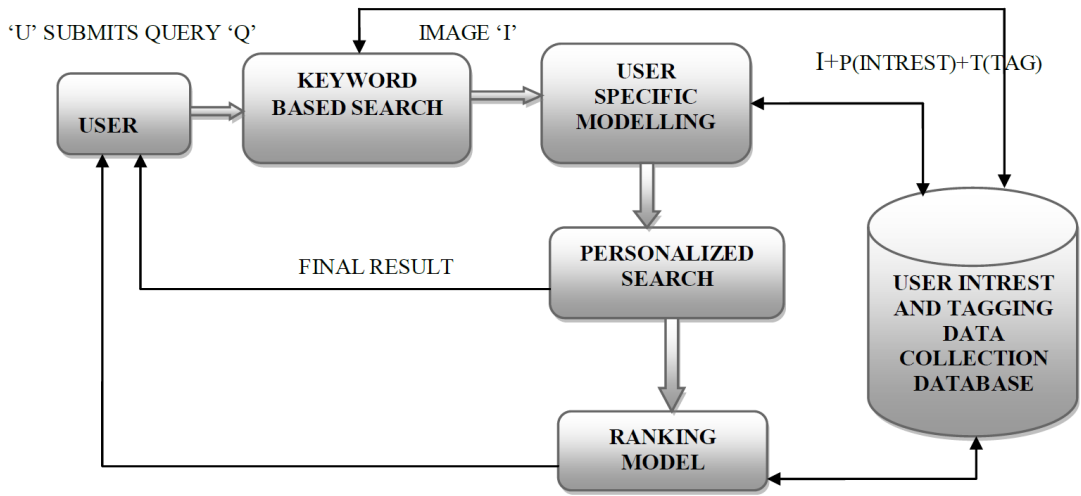

Figure 16: Proposed three-step personalized social image retrieval system [13].

Nowadays, web users create a huge amount of data and a massive quantity of associated metadata. These metadata include tags from images uploaded and shared through social media networks. This vast amount of image metadata can be utilized for personalized image retrieval from social networking sites. However, due to the volume of data, this can be a very difficult task. In [72], the authors proposed a framework that exploits the social activities of users for personalized image searches. Image annotations and interest-based group participation of users are included in these activities. To achieve the expected outcomes, a final rank list is generated by combining query relevance and user preferences.

### 5.3 Personal Photo Curation

With the rise of digital photography through smartphones and other mobile devices and the rapid growth in local and cloud storage capacity, users are now able to have huge personal image collections. This extended image collection has become quite difficult to manage and organize due to the sheer volume and density of the images. In this regard, photo curation has emerged as an important research and application topic. Digital photo curation has been defined as a summation of activities, including deciding which photos to keep, and determining the format, structure, and representation for preservation and display of photos [35].

Photo curation applications can help users to organize their personal image collections and retrieve relevant images easily from these personal repositories. Several research works have explored different aspects of photo curation practices and challenges. Digital photography practices have changed significantly due to advancements in image capturing and media content sharing technologies [24, 43, 67, 80]. In [67], the authors developed a PC-based software app called Shoebox for investigating the digital image collection organizing and browsing behavior of the users. Although this study reported that the browsing facility of digital files provides ease in organization, some other studies found that the huge amount of digital data also makes it very challenging for people to manage and retrieve photos [10, 43, 85]. In [24], the authors developed the PhotoWare system by identifying the requirements for tools to support digital photography. The activities performed by users before and after sharing images were analyzed using the PhotoWork model in [43], which could be used to provide a better analysis of the overflow for the designers. Based on the PhotoWork model, other researchers developed the PhotoUse model [14], which provides a more comprehensive perspective for designing solutions, including cumbersome image tasks such as curation activities. The PhotoUse model categorized photo curation activities into four types of tasks:

- Organizing photos (tag, name, categorize, caption, archive, move, delete photos)
- Triaging photos (evaluate and select photos to share, decorate, or present)
- Managing photos (fill up, download, upload, backup)
- Editing photos (crop, combine, correct, change, retouch photos)

A comprehensive study on curation activities is provided in [92], which identifies design opportunities for future applications to help in managing and organizing photo collections of users. The study showed that, in most cases, an external trigger, such as a shortage of storage space on smartphones is needed to motivate users to perform such curation operations as deleting photos. In most cases, users prefer to use built-in camera rolls for curation activities.

Curation activities can help in organizing, managing photos, and retrieving relevant photos for several tasks, including personal reminiscing, collaborative remembering, or presentation for oneself. In [91], a voting-based photo curation application called Dilemma was developed to coax users into organizing and curating their digital image collections on smartphones. Deleting photos is one of the most frequently performed curation operations, and this operation was addressed using a voting scheme in this work. In [54], a location-based proactive app called Reveal was developed to help users in reminiscing activities. The authors outlined the impact of such tools for reminiscing and emphasized how this type of application can help in supporting effective management and curation of personal photo collections.

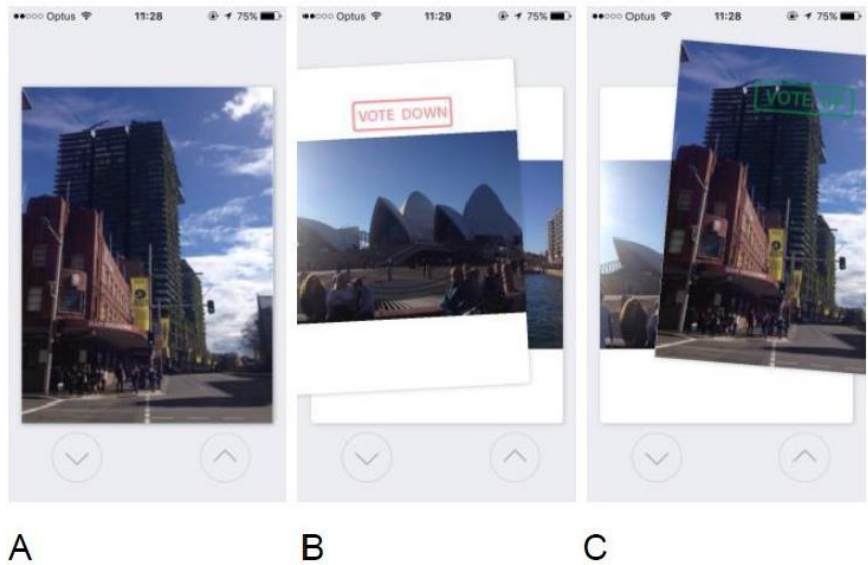
Figure 17: The dilemma voting application with upvote (keep) and downvote (delete) options [92].

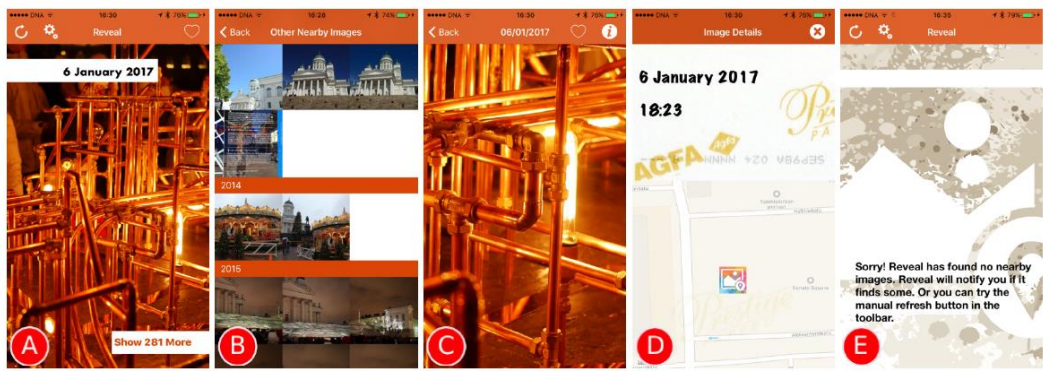
Figure 18: Different views of the Reveal app [54].

With the emergence of human-computer interaction (HCI)-based research, image retrieval research works have also been emphasizing personal image retrieval more and more. There are many unexplored avenues in this domain that need to be explored in the future.

## 6  CONCLUSION

Image retrieval is a challenging research topic with many diverse sub-domains. We have focused on a specific sub-domain of image retrieval called personal image retrieval. The key ideas of personal image retrieval have been explored in this report. Starting with the basics of image retrieval, back-end and front-end research in this domain have been explored. Personalized image retrieval and photo curation have also been analyzed. The overall analysis of all concepts and techniques related to personal image retrieval helps to get a clear picture of what has been done in this research domain and to understand future research directions.

Here we present summarized tabular presentations of all the major concepts discussed in the previous chapters:

Table 2: Back-end Research in Image Retrieval.

| Concepts | Key Ideas | Strengths | Limitations |
|---|---|---|---|
| Content-based Image Retrieval (CBIR) | Retrieves images based on the content present in the images | Provides effective results in general | Performance suffers due to semantic gap |
| Relevance Feedback (RF) | Utilizes user feedback to reduce the semantic gap and enhance retrieval performance | Performs better than traditional CBIR systems | Puts heavy workload on the user |
| Automatic Image Annotation (AIA) | Automates the image annotation process | Makes image retrieval easier and reduces manual annotation time and cost | Demands more computational resources |

Table 3: Front-end Research in Image Retrieval.

| Concepts | Key Ideas | Strengths | Limitations |
|---|---|---|---|
| Similarity-based Visualization | Arranges images based on their visual similarities | Provides effective methods for generating visualizations of large image repositories | Sometimes the mapping generates representations that are too small or congested |
| Interactive Image Retrieval | Providing a scalable visual interface for the users | Creates scopes for more interactive image browsing and navigation | Users need to be initially trained to use the systems |
| Smartphone Applications | Smart device-oriented development and configuration | Ease of use and access for general smart device users | Some applications need to be purchased for use |

Table 4: Personal Image Retrieval and Photo Curation.

| Concepts | Key Ideas | Strengths | Limitations |
|---|---|---|---|
| Human Factors in Image Retrieval | Explores the human factors that have crucial impacts on image retrieval performances | Suggests crucial considerations for developing user-oriented image retrieval systems | Several factors involved could make the design considerations complicated |
| Personalized Image Retrieval | Developing user-oriented image retrieval systems | Provides for more efficient and personalized retrieval results | User preferences can be very subjective thus making the system complex with too many considerations |
| Personal Photo Curation | Organizing and curating personal photo collections | Organized collections help make retrieval easier | Difficult to motivate users to perform photo curation operation |

Existing systems and applications for image retrieval have different kinds of limitations. Most of these applications do not take social media into account. A software/application is needed for retrieving images from the overall image database, which can include device storage, social media, etc.

Also, the performance of an image retrieval system is very subjective as it would differ immensely based on the choice, preference, and selection of each individual. As a result, it is crucial to incorporate real-time human participation to evaluate such systems.

**REFERENCES**

[1] How Many Photos Will Be Taken in 2020? Retrieved March 10, 2021 from https://focus.mylio.com/tech-today/how-manyphotos-will-be-taken-in-2020
[2] Organizing Your Unwieldy Photo Collection Is Easier Than You Think. Retrieved March 20, 2021 from https://www.nytimes.com/2019/04/10/technology/personaltech/photo-organizing-apps.html
[3] Personalized Image Retrieval in social media based on an Optimal Relevance Feedback Algorithm. Retrieved March 22, 2021 from

http://www.wseas.us/e-library/conferences/2013/Vouliagmeni/CCC/CCC-17.pdf

[4] The best photo organizer apps in 2020. Retrieved March 23, 2021 from https://www.tomsguide.com/best-picks/best-photoorganizer-apps

[5] Heelah Alraqibah, Ouiem Bchir, and Mohamed Maher Ben Ismail. X-ray image retrieval system based on visual feature discrimination. volume 9159, 04 2014.

[6] Heelah A. Alraqibah, M. Ismail, and Ouiem Bchir. Empirical comparison of visual descriptors for content based x-ray image retrieval. In ICISP, 2014.

[7] Xavier Anguera, Nuria Oliver, and Mauro Cherubini. Multimodal and mobile personal image retrieval: A user study. 01 2008.

[8] C. R. Anushma. Multimodal image search system on mobile device. International Journal of Computer Organization Trends, 4(2), 2014.

[9] Mohamed Maher Ben Ismail and Hichem Frigui. Unsupervised clustering and feature weighting based on generalized dirichlet mixture modeling. Information Sciences, 274:35–54, 03 2014.

[10] Ofer Bergman, Simon Tucker, Ruth Beyth-Marom, Edward Cutrell, and Steve Whittaker. It's not that important: Demoting personal information of low subjective importance using grayarea. In Proceedings of the SIGCHI Conference on Human Factors in Computing Systems, CHI '09, page 269–278, New York, NY, USA, 2009. Association for Computing Machinery.

[11] James C. Bezdek. Pattern Recognition with Fuzzy Objective Function Algorithms. Kluwer Academic Publishers, USA, 1981.

[12] St´ephane Bissol, Philippe Mulhem, and Yves Chiaramella. Towards personalized image retrieval. 01 2004.

[13] R. Borkar, M. Tamboli, and P. Walnuj. Learn to personalized image search from the photo sharing websites. International Journal of Computer Technology Applications, 4, 2013.

[14] Mendel Broekhuijsen, Elise van den Hoven, and Panos Markopoulos. From photowork to photouse: Exploring personal digital photo activities. Behav. Inf. Technol., 36(7):754–767, July 2017.

[15] P. P. K. Chan, Z. Huang, W. W. Y. Ng, and D. S. Yeung. Dynamic hierarchical semantic network based image retrieval using relevance feedback. In 2011 International Conference on Machine Learning and Cybernetics, volume 4, pages 1746–1751, 2011.

[16] Chaomei Chen, George Gagaudakis, and Paul Rosin. Content-based image visualization. pages 13–18, 02 2000.

[17] Ching-chih Chen, Howard Wactlar, Z. Wang, and Kevin Kiernan. Digital imagery for significant cultural and historical materials an emerging research field bridging people, culture, and technologies. 10 2020.

[18] Qimin Cheng, Qian Zhang, Peng Fu, Conghuan Tu, and Sen Li. A survey and analysis on automatic image annotation. Pattern Recognition, 79, 02 2018.

[19] Philip R. Cohen, Michael Johnston, David McGee, Sharon Oviatt, Jay Pittman, Ira Smith, Liang Chen, and Josh Clow. Quickset: Multimodal interaction for distributed applications. In Proceedings of the Fifth ACM International Conference on Multimedia, MULTIMEDIA '97, page 31–40, New York, NY, USA, 1997. Association for Computing Machinery.

[20] Dacheng Tao, Xiaoou Tang, Xuelong Li, and Xindong Wu. Asymmetric bagging and random subspace for support vector machines-based relevance feedback in image retrieval. IEEE Transactions on Pattern Analysis and Machine Intelligence, 28(7):1088–1099, 2006.

[21] Ritendra Datta, Jia Li, and James Wang. Content-based image retrieval—approaches and trends of the new age. volume 40, pages 253–262, 11 2005.

[22] B. Demir and L. Bruzzone. A novel active learning method in relevance feedback for content based remote sensing image retrieval. IEEE Transactions on Geoscience and Remote Sensing, 53(5):2323–2334, 2015.

[23] Patrick Ehlen and Michael Johnston. Multimodal local search in speak4it. pages 435–436, 01 2011.

[24] David Frohlich, Allan Kuchinsky, Celine Pering, Abbe Don, and Steven Ariss. Requirements for photoware. In Proceedings of the 2002 ACM Conference on Computer Supported Cooperative Work, CSCW '02, page 166–175, New York, NY, USA, 2002. Association for Computing Machinery.

[25] Manali Gaikwad and Orland Hoeber. An interactive image retrieval approach to searching for images on social media. pages 173–181, 03 2019.

[26] Bin Gao, Tie-Yan Liu, Tao Qin, Xin Zheng, QianSheng Cheng, and Wei-Ying Ma. Web image clustering by consistent utilization of visual features and surrounding texts. pages 112–121, 01 2005.

[27] M. Rami Ghorab, Dong Zhou, Alexander O'Connor, and Vincent Wade. Personalised information retrieval: Survey and classification. User Modeling and User-Adapted Interaction, 23, 08 2013.

[28] T. Gokaramaiah, P. Viswanath, and B. E. Reddy. A novel shape based hierarchical retrieval system for 2d images. In 2010 International Conference on Advances in Recent Technologies in Communication and Computing, pages 10–14, 2010.

[29] Yunchao Gong, Qifa Ke, Michael Isard, and Svetlana Lazebnik. A multi-view embedding space for modeling internet images, tags, and their semantics. International Journal of Computer Vision, 106, 12 2012.

[30] A. Grigorova, F. G. B. De Natale, C. Dagli, and T. S. Huang. Content-based image retrieval by feature adaptation and relevance feedback. IEEE Transactions on Multimedia, 9(6):1183–1192, 2007.

[31] Choochart Haruechaiyasak and Chaianun Damrongrat. Improving social tag-based image retrieval with cbir technique. In Gobinda Chowdhury, Chris Koo, and Jane Hunter, editors, The Role of Digital Libraries in a Time of Global Change, pages 212–215, Berlin, Heidelberg, 2010. Springer Berlin Heidelberg.

[32] Y. He, L. Yang, Y. Zhang, X. Wu, and Y. Zhang. The binary image retrieval based on the improved shape context. In 2014 7th International Congress on Image and Signal Processing, pages 452–456, 2014.

[33] Yasuhide Mori Hironobu, Hironobu Takahashi, and Ryuichi Oka. Image-to-word transformation based on dividing and vector quantizing

images with words. In in Boltzmann machines", Neural Networks, page 405409, 1999.

[34] S. C. H. Hoi, M. R. Lyu, and R. Jin. A unified log-based relevance feedback scheme for image retrieval. IEEE Transactions on Knowledge and Data Engineering, 18(4):509–524, 2006.

[35] Nancy Van House and Elizabeth F. Churchill. Technologies of memory: Key issues and critical perspectives. Memory Studies, 1(3):295–310, 2008.

[36] Mohamed Maher Ben Ismail. A survey on content-based image retrieval. International Journal of Advanced Computer Science and Applications, 8, 01 2017.

[37] Alejandro Jaimes. Human factors in automatic image retrieval system design and evaluation. In Simone Santini, Raimondo Schettini, and Theo Gevers, editors, Internet Imaging VII, volume 6061, pages 25 – 33. International Society for Optics and Photonics, SPIE, 2006.

[38] A. K. Jain, M. N. Murty, and P. J. Flynn. Data clustering: A review. ACM Comput. Surv., 31(3):264–323, September 1999.

[39] Zhenyan Ji, Weina Yao, Huaiyu Pi, Wei Lu, Jing He, and Haishuai Wang. A survey of personalised image retrieval and recommendation. National Conference of Theoretical Computer Science, pages 233–247, 10 2017.

[40] Jingzhang, Lizhuo, Lansunshen, and Linhe. A personalized image retrieval based on user interest model. International Journal of Pattern Recognition and Artificial Intelligence, 24, 11 2011.

[41] A. Kannan, V. Mohan, and Neelamegam Anbazhagan. An effective method of image retrieval using image mining techniques. The International journal of Multimedia Its Applications, 2, 12 2010.

[42] Y. Kim, K. Lee, K. S. Choi, J. Yoo, P. Rhee, and Y. Park. Personalized image retrieval with user's preference model. In Other Conferences, 1998.

[43] David Kirk, Abigail Sellen, Carsten Rother, and Ken Wood. Understanding photowork. In Proceedings of the SIGCHI Conference on Human Factors in Computing Systems, CHI '06, page 761–770, New York, NY, USA, 2006. Association for Computing Machinery.

[44] Yanir Kleiman, Joel Lanir, Dov Danon, Yasmin Felberbaum, and Daniel Cohen-Or. Dynamicmaps: Similarity-based browsing through a massive set of images. In Proceedings of the 33rd Annual ACM Conference on Human Factors in Computing Systems, CHI '15, page 995–1004, New York, NY, USA, 2015. Association for Computing Machinery.

[45] Adriana Kovashka and Kristen Grauman. Attribute pivots for guiding relevance feedback in image search. pages 297–304, 12 2013.

[46] P. Kumar, S. Kumar, A. Nimbalkar, and S. Pandey. Multimodal image retrieval on mobile devices. International Journal of Innovative Science, Engineering Technology, 2(3), 2015

[47] Ting-Sheng Lai and John Tait. Chroma (demonstration abstract): A content-based image retrieval system. In Proceedings of the 22nd Annual International ACM SIGIR Conference on Research and Development in Information Retrieval, SIGIR '99, page 324, New York, NY, USA, 1999. Association for Computing Machinery.

[48] Afshan Latif, Aqsa Rasheed, Umer Sajid, Ahmed Jameel, Nouman Ali, Naeem Iqbal Ratyal, Bushra Zafar, Saadat Dar, Muhammad Sajid, and Tehmina Khalil. Content-based image retrieval and feature extraction: A comprehensive review. Mathematical Problems in Engineering, 2019, 08 2019.

[49] Leiquan Wang, Zhicheng Zhao, Fei Su, and Weichen Sun. Content-based social image retrieval with context regularization. In 2014 IEEE International Conference on Multimedia and Expo Workshops (ICMEW), pages 1–6, 2014.

[50] S. Li, M. Lee, and C. Pun. Complex zernike moments features for shape-based image retrieval. IEEE Transactions on Systems, Man, and Cybernetics - Part A: Systems and Humans, 39(1):227–237, 2009.

[51] W. Y. Ma and B. S. Manjunath. Netra: a toolbox for navigating large image databases. In Proceedings of International Conference on Image Processing, volume 1, pages 568–571 vol.1, 1997.

[52] Z. Ma, F. Nie, Y. Yang, J. R. R. Uijlings, and N. Sebe. Web image annotation via subspacesparsity collaborated feature selection. IEEE Transactions on Multimedia, 14(4):1021–1030, 2012.

[53] J. Macqueen. Some methods for classification and analysis of multivariate observations. In In 5-th Berkeley Symposium on Mathematical Statistics and Probability, pages 281–297, 1967.

[54] David K. McGookin. Reveal: Investigating proactive location-based reminiscing with personal digital photo repositories. Proceedings of the 2019 CHI Conference on Human Factors in Computing Systems, 2019.

[55] Ra Mojsilovi, Jos Gomes, and Bernice Rogowitz. Isee: Perceptual features for image library navigation. Proceedings of SPIE - The International Society for Optical Engineering, 4662, 05 2002.

[56] Nils Murrugarra-Llerena and Adriana Kovashka. Cross-modality personalization for retrieval. 2019 IEEE/CVF Conference on Computer Vision and Pattern Recognition (CVPR), pages 6422–6431, 2019.

[57] N. Nacereddine, S. Tabbone, D. Ziou, and L. Hamami. Shape-based image retrieval using a new descriptor based on the radon and wavelet transforms. In 2010 20th International Conference on Pattern Recognition, pages 1997–2000, 2010.

[58] G. Nguyen and M. Worring. Similarity based visualization of image collections. 2005.

[59] Nhu-Van Nguyen, A. Boucher, J. Ogier, and S. Tabbone. Cluster-based relevance feedback for cbir: a combination of query point movement and query expansion. Journal of Ambient Intelligence and Humanized Computing, 3:281–292, 2012.

[60] J´ulia Oliveira, Arnaldo Ara´ujo, and Thomas Deserno. Content-based image retrieval applied to bi-rads tissue classification in screening mammography. World journal of radiology, 3:24– 31, 01 2011.

[61] P. B. Patil and M. B. Kokare. Content based image retrieval with relevance feedback using riemannian manifolds. In 2014 Fifth International Conference on Signal and Image Processing, pages 26–29, 2014.

[62] W. Pedrycz, V. Loia, and S. Senatore. Fuzzy clustering with viewpoints. IEEE Transactions on Fuzzy Systems, 18(2):274–284, 2010.

[63] W. Plant and G. Schaefer. Navigation and browsing of image databases. In 2009 International Conference of Soft Computing and Pattern Recognition, pages 750–755, 2009.

[64] W. Plant and G. Schaefer. Visualising image databases. In 2009 IEEE International Workshop on Multimedia Signal Processing, pages 1–6, 2009.

[65] Adrian Popescu and Gregory Grefenstette. Social media driven image retrieval. page 33, 01 2011.

[66] B. E. Prasad, A. Gupta, H. D. Toong, and S. E. Madnick. A microcomputer-based image database management system. IEEE Transactions on Industrial Electronics, IE-34(1):83–88, 1987.

[67] Kerry Rodden and Kenneth R. Wood. How do people manage their digital photographs? In Proceedings of the SIGCHI Conference on Human Factors in Computing Systems, CHI '03, page 409–416, New York, NY, USA, 2003. Association for Computing Machinery.

[68] S. Santini, A. Gupta, and R. Jain. Emergent semantics through interaction in image databases. IEEE Transactions on Knowledge and Data Engineering, 13(3):337–351, 2001.

[69] G. Schaefer. Interactive browsing of large image databases. In 2016 Sixth International Conference on Digital Information Processing and Communications (ICDIPC), pages 168–170, 2016.

[70] Uzma Sharif, Zahid Mehmood, Toqeer Mahmood, Dr Javid, Amjad Rehman, and Tanzila Saba. Scene analysis and search using local features and support vector machine for effective content-based image retrieval. Artificial Intelligence Review, 06 2018.

[71] Saurabh Sharma, Vishal Gupta, and Mamta Juneja. A survey of image data indexing techniques. Artificial Intelligence. Review, 52(2):1189–1266, August 2019.

[72] Dr. Amit Sinhal. Personalized image search from photo sharing websites using ranking based tensor factorization model (rmtf). International Journal of Advanced Research in Computer Science and Software Engineering, 3:652–656, 08 2013.

[73] Noah Snavely, Steven Seitz, and Richard Szeliski. Photo tourism: exploring photo collections in 3d. acm trans graph 25(3):835-846. ACM Trans. Graph., 25:835–846, 07 2006.

[74] G. Strong and Minglun Gong. Browsing a large collection of community photos based on similarity on gpu. In ISVC, 2008.

[75] Zhong Su, Hongjiang Zhang, Stan Li, and Shaoping Ma. Relevance feedback in contentbased image retrieval: Bayesian framework, feature subspaces, and progressive learning. IEEE transactions on image processing: a publication of the IEEE Signal Processing Society, 12:924–37, 02 2003.

[76] Nicolae Suditu and Fran¸cois Fleuret. Iterative relevance feedback with adaptive exploration/exploitation trade-off. In Proceedings of the 21st ACM International Conference on Information and Knowledge Management, CIKM '12, page 1323–1331, New York, NY, USA, 2012. Association for Computing Machinery.

[77] Eric Tellez, Edgar Ch´avez, and Jos´e Ortiz-Bejar. Scalable proximity indexing with the list of clusters. 11 2014.

[78] Ricardo S. Torres, Celmar G. Silva, Claudia B. Medeiros, and Heloisa V. Rocha. Visual structures for image browsing. In Proceedings of the Twelfth International Conference on Information and Knowledge Management, CIKM '03, page 49–55, New York, NY, USA, 2003. Association for Computing Machinery.

[79] Shereena V B and Julie David. Content based image retrieval: Classification using neural networks. The International journal of Multimedia Its Applications, 6:31–44, 10 2014.

[80] Nancy A. Van House. Collocated photo sharing, story-telling, and the performance of self. International Journal of Human-Computer Studies, 67(12):1073 – 1086, 2009. Collocated Social Practices Surrounding Photos.

[81] M. Singh W. Ren and C. Singh. Image retrieval using spatial context. In Ninth International Workshop on Systems, Signals and Image Processing, 2002.

[82] Chaoli Wang, John P Reese, Huan Zhang, Jun Tao, Yi Gu, Jun Ma, and Robert J Nemiroff. Similarity-based visualization of large image collections. Information Visualization, 14(3):183–203, 2015.

[83] J. Wang, W. Liu, S. Kumar, and S. Chang. Learning to hash for indexing big data—a survey. Proceedings of the IEEE, 104(1):34–57, 2016.

[84] James Wang, Gio Wiederhold, Oscar Firschein, and Sha Wei. Content-based image indexing and searching using daubechies' wavelets. International Journal on Digital Libraries, 1:311–328, 03 1998.

[85] Steve Whittaker, Ofer Bergman, and Paul Clough. Easy on that trigger dad: A study of long term family photo retrieval. Personal and Ubiquitous Computing, 14:31–43, 01 2010.

[86] X. Xie, L. Lu, M. Jia, H. Li, F. Seide, and W. Ma. Mobile search with multimodal queries. Proceedings of the IEEE, 96(4):589–601, 2008.

[87] Jing Zhang, Lansun Shen, and David Dagan Feng Feng. A personalized image retrieval based on visual perception. Journal of Electronics (China), 25:129–133, 01 2008.

[88] Lei Zhang and Yong Rui. Image search-from thousands to billions in 20 years. ACM Transactions on Multimedia Computing, Communications, and Applications, 9:1–20, 10 2013.

[89] Xiaofeng Zhang, Kwok-Wai Cheung, and Chun-hung Li. Learning latent variable models from distributed and abstracted data. Inf. Sci., 181:2964–2988, 07 2011.

[90] Changyun Zhu, Kun Li, Qin Lv, li Shang, and Robert Dick. iscope: Personalized multimodality image search for mobile devices. pages 277–290, 01 2009.

[91] Xenia Z¨urn, Mendel Broekhuijsen, Dom´enique van Gennip, Saskia Bakker, Annemarie Zijlema, and Elise van den Hoven. Stimulating photo curation on smartphones. In Proceedings of the 2019 Conference on Human Information Interaction and Retrieval, CHIIR '19, page 255–259,

New York, NY, USA, 2019. Association for Computing Machinery.

[92] Xenia Z̈urn, Koen Damen, Fabienne van Leiden, Mendel Broekhuijsen, and Panos Markopoulos. Photo curation practices on smartphones. In Adrian David Cheok, Masahiko Inami, and Teresa Romão, editors, Advances in Computer Entertainment Technology, pages 406–414, Cham, 2018. Springer International Publishing.

[93] Anjana Gosain and Sonika Dahiya. Performance analysis of various fuzzy clustering algorithms: A review. Proceedings of International Conference on Communication, Computing and Virtualization (ICCCV) 2016.